\documentclass[12pt,preprint]{aastex}

\slugcomment{Accepted by the PASP} \shorttitle{Resolved and
Unresolved Interferometer Calibrators} \shortauthors{van Belle and
van Belle}

\begin{document}
\title{Establishing Visible Interferometer System Responses:\\
Resolved and Unresolved Calibrators}
\author{Gerard T. van Belle}
\affil{Michelson Science Center, California Institute of Technology,
      Pasadena, CA 91125} \email{gerard@ipac.caltech.edu}
\and
\author{Gerald van Belle}
\affil{Department of Biostatistics,
University of Washington, Seattle, WA 98195-7232}
\email{vanbelle@seanet.com}

\begin{abstract}
The propagation of errors through the uniform disk visibility
function is examined.  Implications of those errors upon measures of
absolute visibility through optical and near-infrared
interferometers are considered within the context of using
calibration stars to establish system visibilities for these
instruments. We suggest a simple ratio test to establish empirically
whether or not the measured visibilities produced by such an
instrument are relative (errors dominated by calibrator angular size prediction error)
or absolute (errors dominated by measurement error).
\end{abstract}


\section{Introduction}

Visible and near-infrared interferometers are powerful tools for
measuring the minute angular sizes of nearby stars. However,
establishing absolute system responses in the presence of
atmospheric turbulence and instrument imperfections is a challenging
proposition that requires careful attention to detail when
constructing an observational approach.

For two-element interferometers that are now commonly in use, the
principal measured quantity is the visibility, $V$, which
is simply a characterization of the contrast found in
the observed interference fringe and can range from 0 to 1.
In practice, interferometers
that lock onto and track fringes through temporally modulating servo
loops tend to measure $V^2$ rather than just $V$.
A detailed discussion of fringe visibility estimators can
be found in \citet{colav1999}.

For individual stars, the observed $V^2$ will decrease from unity
as the source becomes resolved to the instrument, and also
as the response of the instrument and atmosphere through which it observes
departs for an idealized system.
A common approach to account for the system response (a
combination of the atmospheric and instrumental responses) is to
interleave observations of calibration stars with observations of
the star of interest.  If the system visibility ${V^2}_{system}$ is established with calibration
sources, a target star's absolute $V^2$ is then easily derived from
the measured $V^2$:
\begin{equation}\label{eqn_measSys}
{V^2}_{normalized}(target) = { {V^2}_{measured}(target) \over {V^2}_{system} }
\end{equation}
Calibration sources are stars which have
some sort of {\it a priori} knowledge of their angular size, and as
a result, their expected values for ${V^2}_{predicted}$ can be predicted.  Thus, from
the calibrator's measured values for ${V^2}_{measured}$, the
system visibility ${V^2}_{system}$ that characterizes the
atmospheric and instrumental performance degredations is simply:
\begin{equation}\label{eqn_measExp}
{V^2}_{system} = { {V^2}_{measured}(calibrator) \over {V^2}_{predicted}(calibrator) }
\end{equation}
It is important to note that, as is apparent from the practice of using the
system visibility found in Equation \ref{eqn_measExp} in Equation \ref{eqn_measSys}, the
system visibility needs to be constant when observing both the
target star and the calibration star.
This consideration is significant when
such objects are of differing brightness, are located in different
portions of the sky, subject to varying weather conditions, or are widely separated in time. Insofar that no two stars will
be of exactly the same brightness, the instrument will be required to have some
measure of dynamic range in this regard; for the purposes of this investigation,
we will assume all of the relevant data are properly collected within the range
of constant system visibility.

For some of the sources considered for use as calibrators, angular
sizes have actually been measured.  However, for most of the sources
considered as calibrators, some sort of indirect estimation
technique needs to be employed.  These techniques include angular size
estimation from distance and linear size (with the second quantity often
being inferred from some proxy such as spectral type), blackbody
fitting, R-I and V-K proxies \citep{mozur1991,diben1993,vanbe1999}, and
spectrophotometric fits
\citep{black1994,cohen1996}.

For unbiased results, it is preferred to utilize calibration sources that are `unresolved'
to the interferometer.  A source is considered to be unresolved when the
errors in ${V^2}_{system}$ are dominated by the measurement error
and not the prediction error $\left( (\sigma_{V^2})_{measured} > (\sigma_{V^2})_{predicted} \right)$.
For such a calibrator, biases in the angular size
estimation technique - known or unknown - are masked by the
measurement process.  This is due the interferometer's insensitivity
to the unresolved calibrator's angular size, and as such, this
technique is insensitive to estimation technique biases.
In the limit that the instrument performance is linear between
target and calibration sources, the resultant
calibration thus is considered to be an {\it absolute calibration}
of the system visibility ${V^2}_{system}$.
This approach is documented in the literature for many
interferometers, including the Mark III \citep{mozur1991}, IRMA \citep{dyck1993},
IOTA \citep{dyck1996},
and PTI \citep{vanbe1999b}.  Additionally, as we shall see in \S \ref{sec_Bias}, this approach
avoids the regime where the separate Taylor series bias due to non-linearity in the error propagation technique
becomes significant as well.

Another approach seen with some regularity is the establishment of
system ${V^2}_{system}$ through use of resolved calibrators. In the case where
instrumental limitations (typically limited sensitivity) preclude
the use of an unresolved calibrator, investigators have utilized
calibration sources that are resolved to establish instrument system
visibilities.  The strength of this approach is that resolved
calibrators are typically associated with stars of greater
brightness, and as a result, greater signal-to-noise is achieved in
observing the calibration sources.

The weakness of this approach is that it establishes only a {\it
relative calibration} for the measurement, and any biases inherent
in the original size estimation of the calibration source
propagate into the final visibility measured for the target source, albeit with
additional uncertainty due to measurement error in ${V^2}_{measured}$.
Relative $V^2$ measures, when properly used, are useful quantities for certain investigations -
for example, the examination of the shape of a rotationally distorted star
\citep{domic2003}
- but are inappropriate to use as absolute values to establish quantities
such as stellar linear size or effective temperature.

\section{The Visibility Function and \\ Angular Size Estimation Bias}\label{sec_visfn_and_errors}
The projections of stellar disks upon the sky are clearly not true
`uniform disks' (see \citet{hajian1998} and references therein), having varying brightness from the center to the
edge of their disks.  However, for most stars, characterization of
them as uniform disks is a reasonable approximation, and one which
will lend itself to a mathematical examination in \S \ref{errprop}.

A uniform disk as viewed by an interferometer exhibits a visibility
function $w$ given by:
\begin{equation}\label{UDeqn}
V(x)^2 = w(x)
    = \left({2 J_1(x) \over x}\right)^2
    = \left({2 J_1({\pi \theta B / \lambda }) \over {\pi \theta B / \lambda }}\right)^2
\end{equation}
where $x$ is the spatial frequency,
and a function of projected baseline $B$, source angular size $\theta$,
and observational wavelength $\lambda$:
$x=\pi \theta B / \lambda$ \citep{airy1835, born1980}.

Since we will be utilizing calibration sources with predicted
angular sizes $\theta$, it is of great utility to examine the impact that
errors (and potentially size estimation biases) have upon our expected values for
calibration source ${V^2}_{predicted}$.

\subsection{Uniform Disk Visibility Error Propagation\label{errprop}}

Since
$w=w(\theta,B,\lambda)$, a routine propagation of errors through Equation \ref{UDeqn} gives:
\begin{equation}\label{propOfErrors}
\sigma_w^2 = \left(\frac{\partial w}{\partial \theta}\right)^2
\sigma_\theta^2
      + \left(\frac{\partial w}{\partial B}\right)^2 \sigma_B^2
      + \left(\frac{\partial w}{\partial \lambda}\right)^2 \sigma_\lambda^2
      + \textrm{cov}(\theta,B,\lambda)
\end{equation}
with the covariance terms for this discussion expected to be zero
(we will reexamine the higher order terms of Equation \ref{propOfErrors} in
\S \ref{sec_Bias}.)  For evaluation of Equation \ref{propOfErrors}, it
is useful to employ the $jinc(x)$
function, which is defined in \citet{brace2000}, and its first derivative given as
\begin{equation}\label{jincFunction}
jinc(x) = {J_1(x) \over x} \textrm{ and } jinc'(x)=-{J_2(x) \over x}.
\end{equation}
Using the chain rule on Equations \ref{UDeqn} and \ref{propOfErrors}
\begin{equation}
\frac{\partial w}{\partial x} = \frac{\partial }{\partial x}
\left[\left({2J_1(x) \over x}\right)^2\right] = \frac{\partial
}{\partial x} 4jinc(x)^2 = 8jinc(x)jinc'(x) = -{8J_1(x) J_2(x) \over
x^2}
\end{equation}
Equation \ref{propOfErrors} can be rewritten as:
\begin{equation}
\sigma_w^2 = \left[{-{8J_1(x)J_2(x) \over x^2}}\right]^2 \left[ {
  \left[ {\pi B \over \lambda} \right]^2 \sigma_\theta^2
+ \left[ {\pi \theta \over \lambda} \right]^2 \sigma_B^2 + \left[
-{\pi B \theta \over \lambda^2} \right]^2 \sigma_\lambda^2 } \right]
\end{equation}
\begin{equation}\label{eqn9}
= \left[{{8J_1(x)J_2(x) \over x}}\right]^2 \left[ {
  \left( {\sigma_\theta \over \theta}\right)^2
+ \left( {\sigma_B      \over B     }\right)^2 + \left(
{\sigma_\lambda\over \lambda}\right)^2 } \right]
\end{equation}
In the limit that the calibrator size prediction fractional errors
dominate ($\sigma_\theta / \theta >> \sigma_B / B, \sigma_\lambda /
\lambda $), we have
\begin{equation}\label{eqn10}
(\sigma_{V^2})_{predicted}  = \sigma_w  =
\left[{{8J_1(x)J_2(x) \over x}}\right]
  \left[ {\sigma_\theta \over \theta} \right]
\end{equation}
This error propagates in quadrature back to our estimate of the
system visibility in Equation \ref{eqn_measExp} along with any
measurement error, $(\sigma_{V^2})_{measured}$.

For consideration of $(\sigma_{V^2})_{measured}$, it is very important
to not only establish the measurement scatter of a single $V^2$ sampling
event, but to empirically establish the night-to-night measurement
error found in $V^2$ measurements.  An excellent example of such
a characterization is the examination of the final $V^2$ residuals
in the binary star fit of $\iota$ Pegasi found in \citet{bod99}.

\subsection{Absolute versus Relative Ratio Test\label{merit_eval}}
We may use Equations \ref{eqn9} and \ref{eqn10} to explore the
impact of calibrator size prediction error $\sigma_\theta$ upon the
system visibility error $(\sigma_{V^2})_{predicted}$.  Our test
case will be as follows: a 330-m baseline with a $\sigma_B=$1-cm
error in its knowledge of projection upon the sky (which will be the
product of geometry knowledge errors and timing errors, but still is
a generous error bar for this term); a $\sigma_\lambda=$0.01$\mu$m
error in the knowledge of the operational wavelength,
$\lambda=2.15\mu$m; and a 5\% prediction error $\sigma_\theta$ in
the angular size estimate for an individual calibrator.  This test
case is fairly representative of the current parameters of interest
for the CHARA Array \citep{tenbrum2005}.

      \begin{figure}
      \plotone{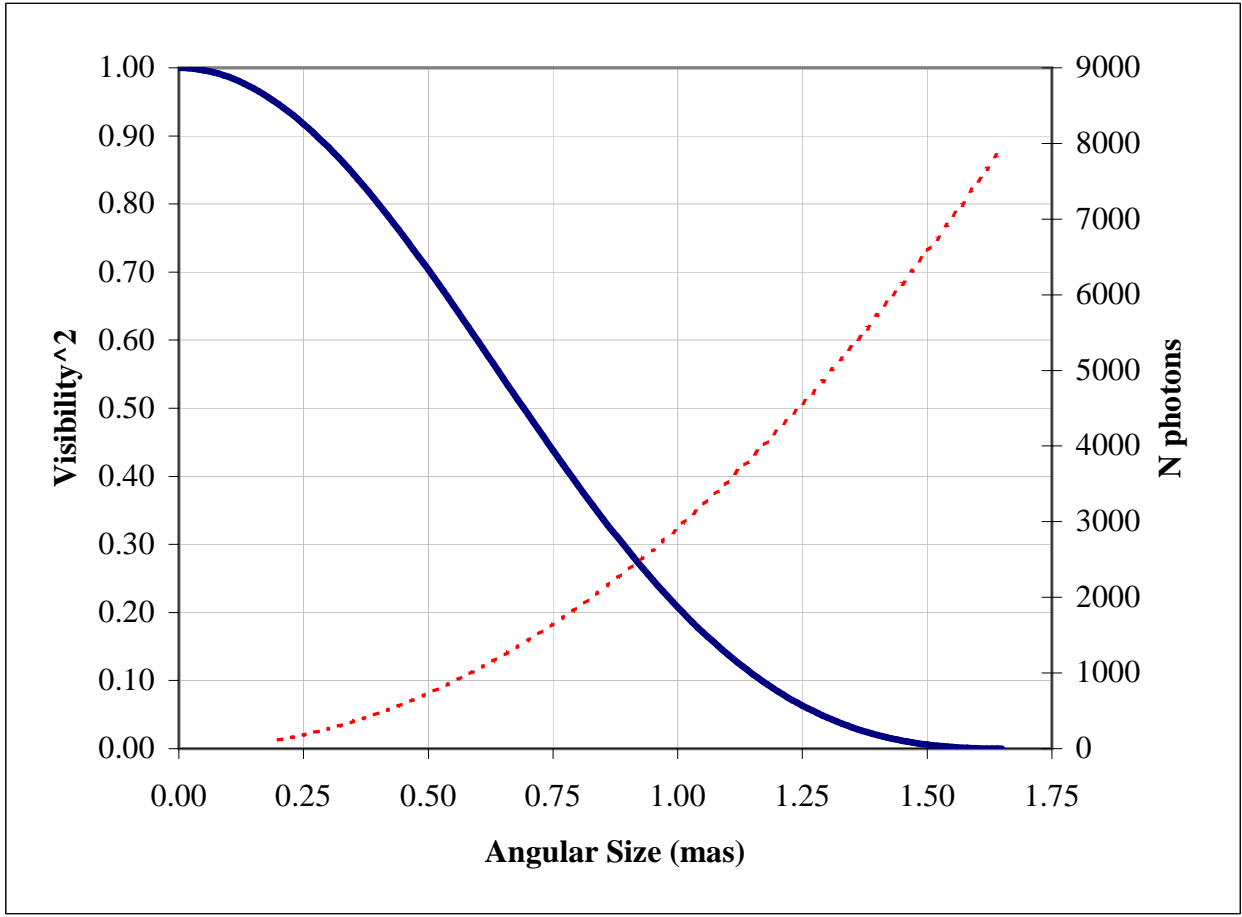}
      \caption{\label{fig1}$V^2$ for uniform disk stars as viewed by the CHARA Array, with a 330-m baseline
      and a
      $K_s$ bandpass.  Also shown on the right vertical axis with a red dotted line
      is detected photon count $N$, assuming a G2V calibration source,
      a 0.001 ms integration
      time per sample, 1-m diameter aperture, and 4\% throughput.}
      \end{figure}

\begin{figure}
      \plotone{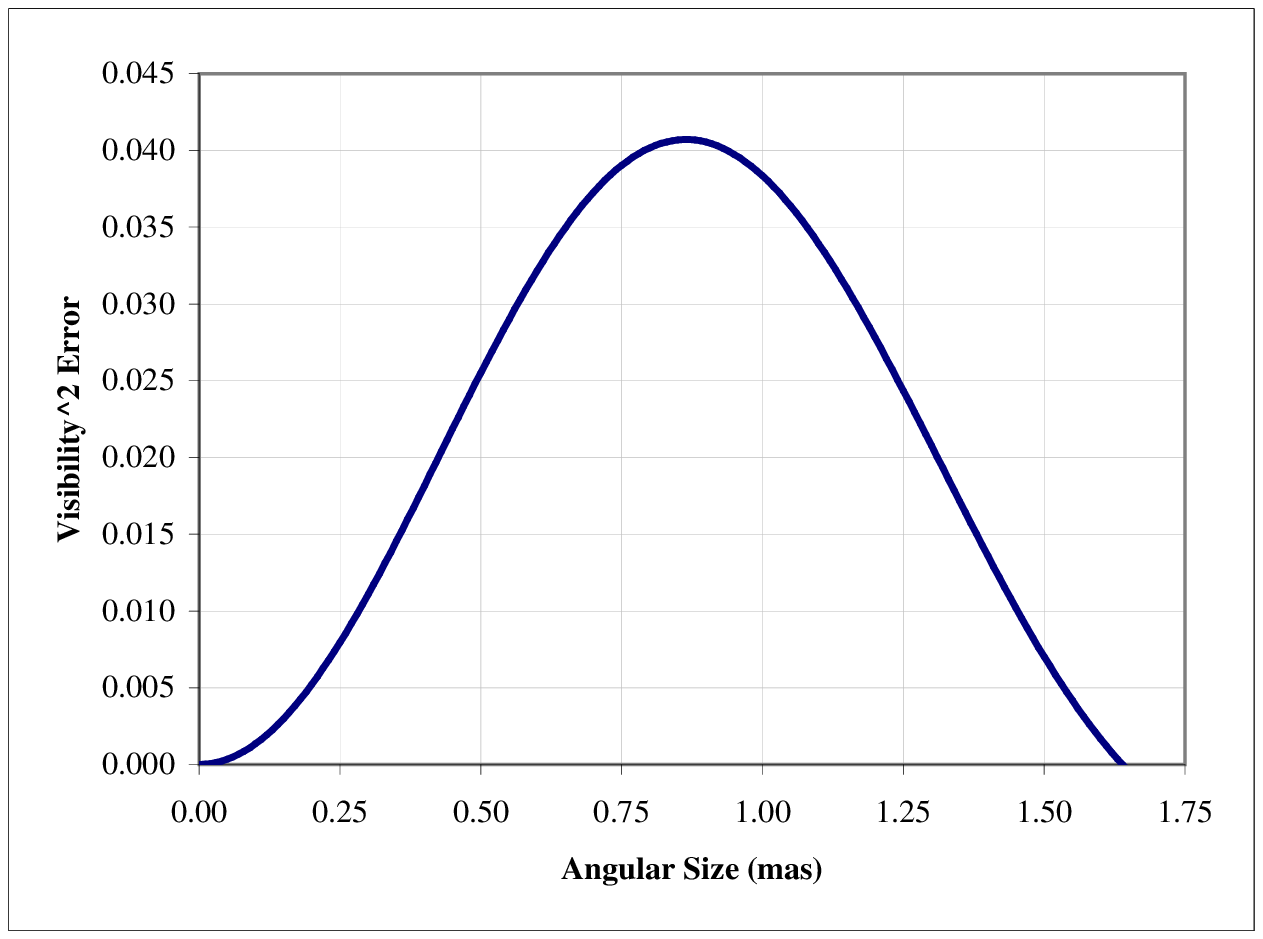}
      \caption{\label{fig2}Calibrator $V^2$ prediction error $(\sigma_{V^2})_{predicted}$
      propagated from an assumed 5\% uncertainty
      in calibrator angular size, {\bf not} accounting for measurement error.}
      \end{figure}

\begin{figure}
      \plotone{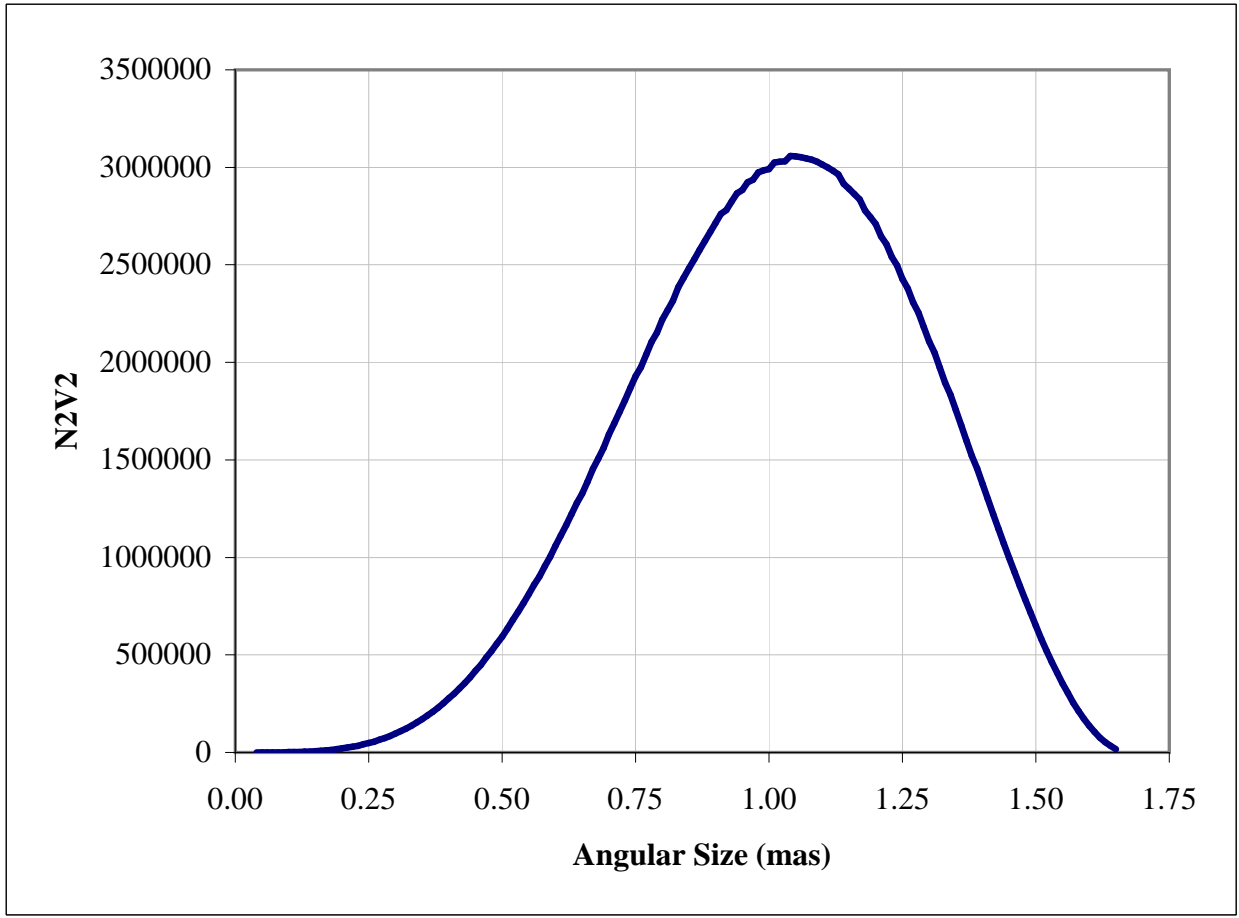}
      \caption{\label{fig3a}Signal-to-noise ($N^2V^2$) for our G2V source as a function of
      calibrator angular size.}
      \end{figure}

For a range of angular sizes, the predicted $V^2$ value for the
calibrator is plotted in Figure 1.  As the star passes 0.690 mas,
already $V^2$ has fallen below 50\%, and drops to zero beyond 1.500
mas.  Also plotted in Figure 1 on the righthand axis is a rough
expectation of number of detected photons $N$ for the CHARA Array for a
G2V star, following the relationship detailed in \citet{vanbe1999}
between $V-K$ and angular size (noting that the estimate of system
throughput may be inaccurate but only scales the results herein).
For our hypothetical G2V star we have $V-K=1.5$ \citep{besse1988}.
The predicted $V^2$ error derived from those values for the three
error terms in Equation \ref{eqn9} is shown in Figure \ref{fig2}.
We may also estimate our signal-to-noise as being proportional to $N^2V^2$ in the read noise
limited regime (the usual operational case for near-infrared
interferometers) \citep{colav1999}, although we note that a similar analysis
we have executed for $NV^2$ gives results similar to those presented in this
section and the next.  (This
latter case corresponds to photon noise-limited operations \citep{mozur1991}
in the low photon limit, as can be the case for visible interferometers.)
A plot of this is shown in
Figure \ref{fig3a}, which is effectively is the product of the
solid line and the dotted line squared seen in Figure \ref{fig1}.

\begin{figure}
      \plotone{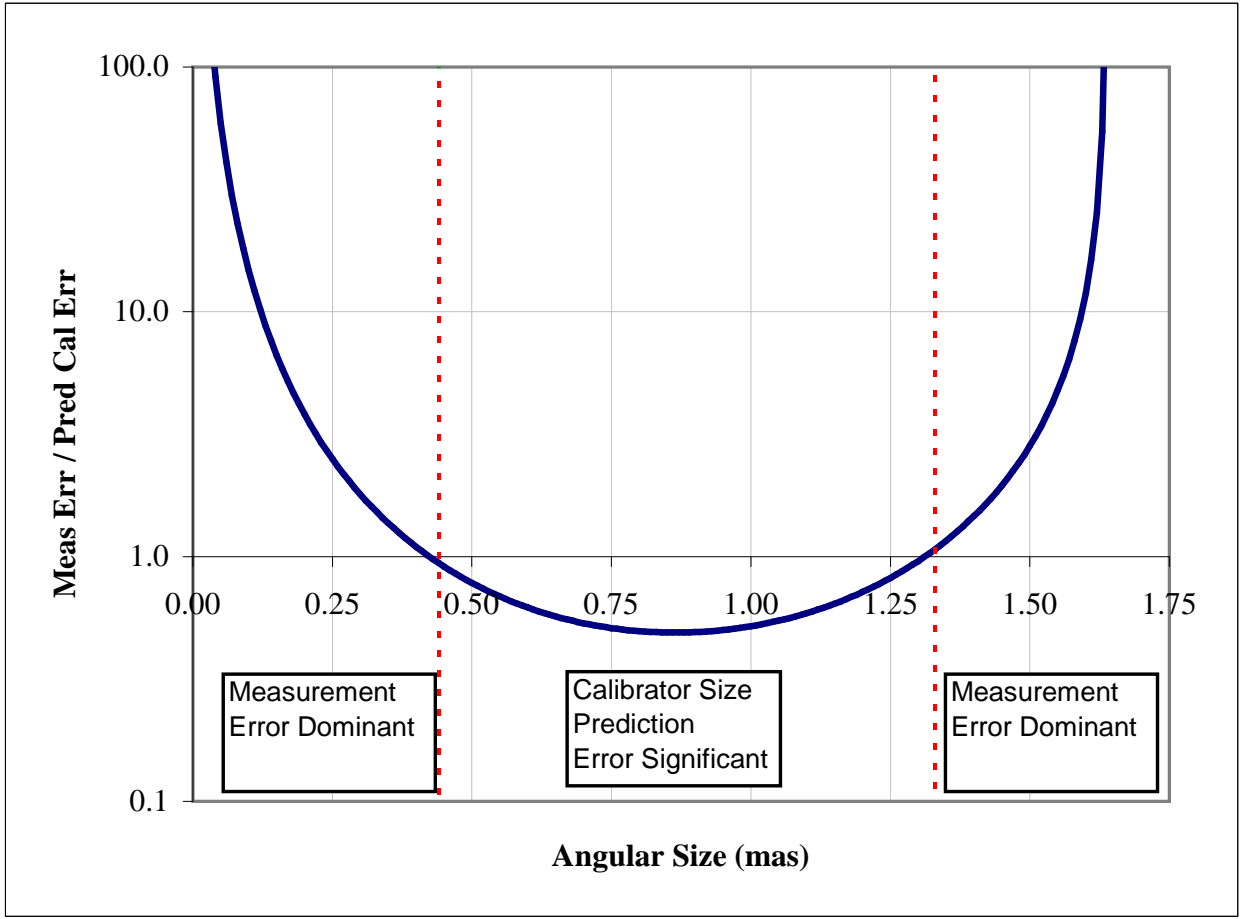}
      \caption{\label{fig5}From Equation \ref{ratioTest}, the ratio of an assumed 2\% $V^2$ measurement error to the $V^2$ error due to
      calibrator size prediction error of 5\%, as a function of expected calibrator size.  Note the
      regime between the red dotted lines, $0.45 < \theta < 1.35$ mas, where the
      $(\sigma_{V^2})_{predicted}$ has a significant impact upon the final errors.}
      \end{figure}

A way to illustrate this point is to examine the ratio $r$ of
$V^2$ measurement error $(\sigma_{V^2})_{measured}$ to
calibrator $V^2$
prediction error $(\sigma_{V^2})_{predicted}$, as is seen in
Figure \ref{fig5}:
\begin{equation}\label{ratioTest}
r = {(\sigma_{V^2})_{measured} \over (\sigma_{V^2})_{predicted}}
\end{equation}
The range of angular sizes where
$r$ dips below 1.0 indicates where $(\sigma_{V^2})_{predicted}$
is a significant, if not the dominant, contribution to the $V^2$
measurement error.
Equation \ref{ratioTest} (with the denominator as provided by Equation
\ref{eqn9} or \ref{eqn10}) is a straightforward indicator of the
interferometer operational regime as determined by the
choice of calibrator: for $r<1$, it is relative, and for $r>1$,
it is absolute.

It is interesting to note that Figure \ref{fig5} indicates
a second regime of absolute calibrator sizes for our example
case, that of the `superresolved' sources in the range of
$\theta > 1.35$ mas.  Simply put, it is in this regime that the
$V^2$ function has once again flattened out (see Figure \ref{fig1})
and uncertainty in $\theta$ does little to impact $V^2_{predicted}$ for
the calibrator.
Unfortunately, it is in this range that the signal-to-noise
is rapidly dropping to zero, as seen already in Figure \ref{fig3a}.
Also, as we will see in \S \ref{sec_Bias}, this regime is problematic
due to bias in the error propagation technique.

\subsection{A Merit Function and its Evaluation}\label{sec_ratio_test}

\begin{figure}
      \plotone{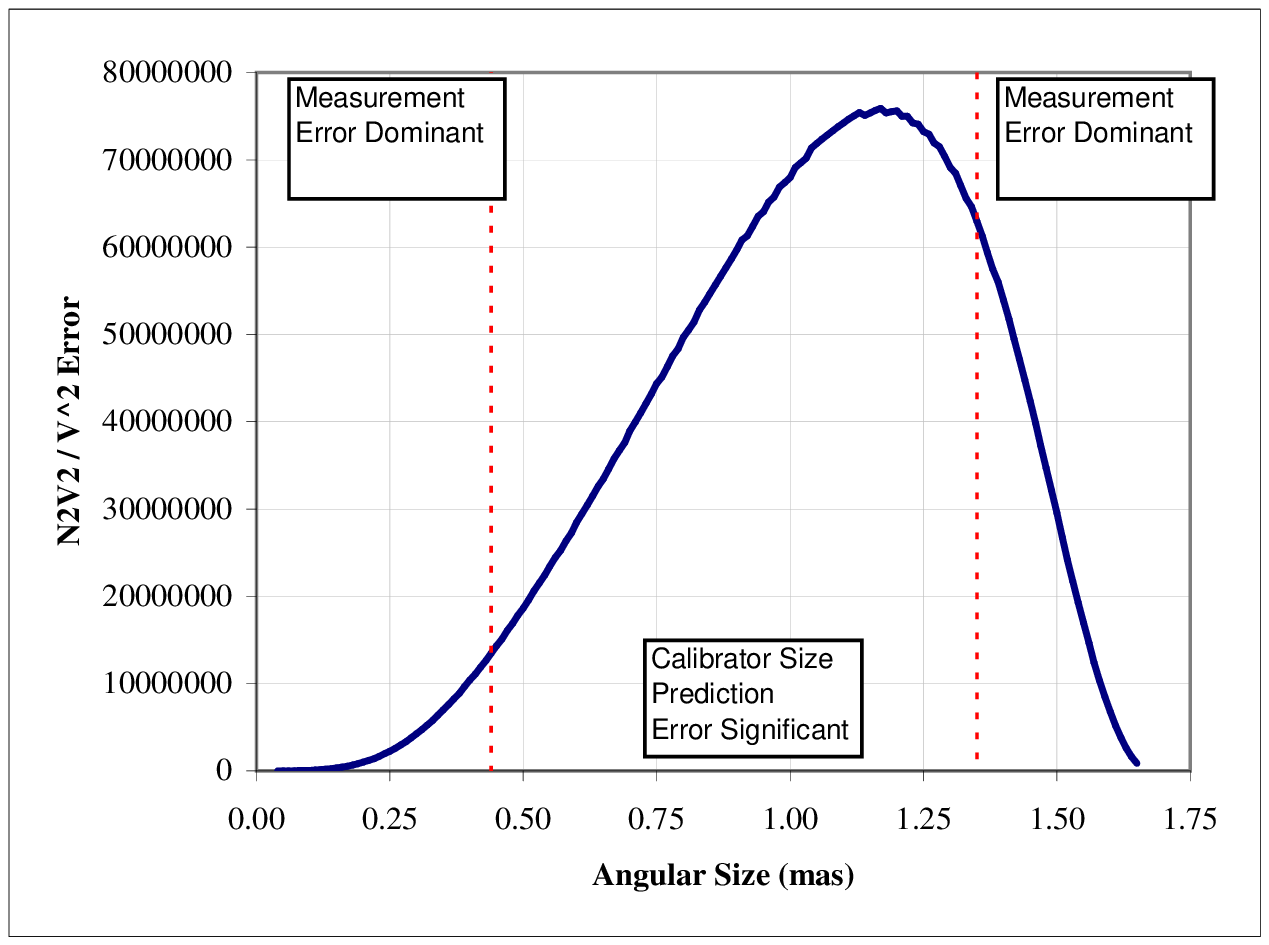}
      \caption{\label{fig4}Full calibrator merit function, $N^2V^2 / \sigma_{V^2}$, propagated from an assumed 5\% uncertainty
      in calibrator angular size, and including an putative 2\% $V^2$ measurement error.  The
      regime to the left of the red dotted line (for this example case, $0<\theta<0.45$ mas) has measurement error
      as the dominant contribution to the merit function.}
      \end{figure}

As a useful metric of `calibrator goodness', we propose a merit
function equal to the ratio of signal-to-noise to system visibility
error:
\begin{equation}\label{meritTest}
m = {N^2V^2 \over \sigma_{V^2}}
\end{equation}
In the real world case, $V^2$ measurement error
$(\sigma_{V^2})_{measured}$ also affects our measures of the system
visibility.  The resultant system visibility error is computed from
the measurement error and the calibrator
$V^2$ prediction error, added in quadrature:
\begin{equation}
\sigma_{V^2}^2 = (\sigma_{V^2})_{measured}^2 + (\sigma_{V^2})_{predicted}^2
\end{equation}
and applied to our merit function.  The merit function incorporating
the measurement error and calibrator size prediction error
is plotted in Figure \ref{fig4}, for the case where
the $V^2$ measurement errors are assumed to be at the 2\% level.

Of interest in our CHARA Array example are the angular sizes beyond $\sim$0.450 mas, where the
merit function changes slope, peaks, and descends. It is at
those angular sizes ($\theta > 0.450$ mas) that
the contribution to the merit
function transitions from $V^2$ measurement error
$(\sigma_{V^2})_{measured}$ being dominant to calibrator $V^2$
prediction error $(\sigma_{V^2})_{predicted}$ becoming significant
and then dominant.  This regime is of particular interest:
if the technique being employed for calibrator $V^2$ prediction is
subject to a systematic size estimation bias due to an imperfection in the
predictive technique, that bias will begin to significantly affect
the inferred $V^2$ values for calibrators in excess of the size,
despite their apparent greater `merit' indicated by Equation \ref{meritTest}.
For calibrators short of this
point, any systematic size estimation biases in the size prediction (and resultant
$V^2$ prediction) will be masked by the calibrator's point-like
nature for the interferometer system.

\subsection{Stellar Angular Size Prediction Bias \\ Example: The Blackbody Case}

If we consider (as is frequently done) stars as blackbodies, we may
fit broad and narrow-band photometry from these objects with a
Planck function, which will result in predictions for the object's
effective temperature, bolometric flux, and angular size.  However,
such an approximation is quite poor and overlooks many subtleties of
stellar atmospheres, such as wavelength-dependent opacities.

\begin{figure}
      \plotone{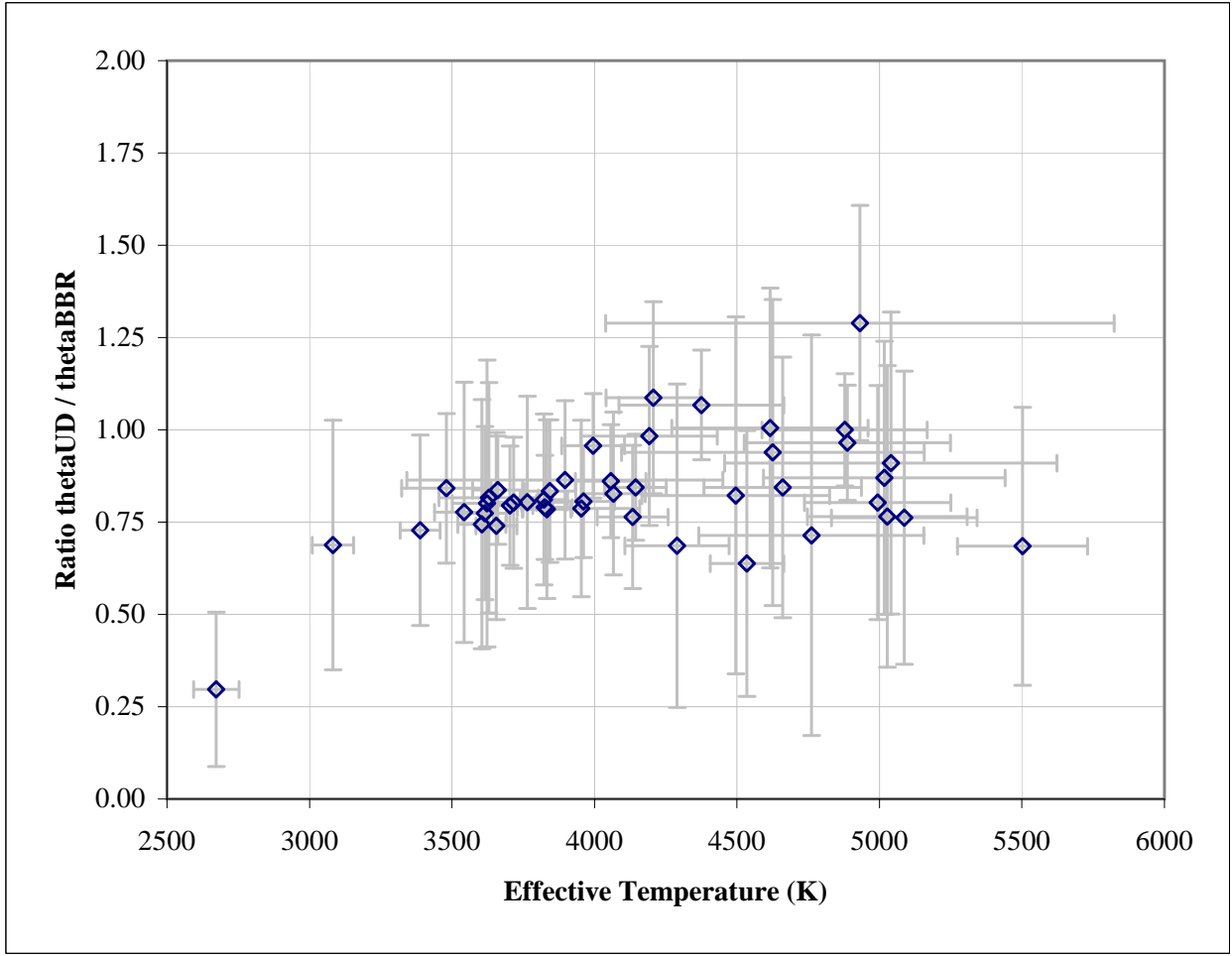}
      \caption{\label{fig0}The ratio of measured angular size to predicted angular
      size (as derived from a blackbody approximation)
      as a function of effective temperature for 48 stars from \citet{vanbe1999b}. }
      \end{figure}

In order to quantify the specifics of this example, a sample of 48
late giant stars from \citet{vanbe1999b} that were
well-characterized photometrically was examined with such an
approximation.  This sample has the benefit of having sizes measured
and presented in \citet{vanbe1999b} for comparison to the results
obtained with the blackbody fit.  A plot of the ratio of measured
angular size to blackbody-derived angular size, as a function of the
effective temperatures established for those stars in the paper, is
shown in Figure \ref{fig0}. Errors in the blackbody angular size were
derived from appropriate iteration of the Planck function within the
errors specified for the photometry.

What is interesting to note in Figure \ref{fig0} is the systematic offset
of the ratios below a line of unity - the blackbody technique
systematically delivers an angular size that is too large relative
to the sizes that have been measured. The errors in that ratio,
propagated from the blackbody and measured angular size errors,
indicate that the ratio of unity is within most of the displayed error bars,
as one would expect, but the general trend (on
the order of $\sim$15-25\%) shows that use of simple blackbody angular
sizes could potentially bias interferometer calibrators.

As an example, if we were to use these sorts of stars in this manner with the
Palomar Testbed Interferometer, we would find that for $K$-band operations
with its 110-m baseline, we would need stars in the $\theta \leq 0.45-58$ mas range
for use as absolute calibrators, given its limiting measurement precision
of $(\sigma_{V^2})_{measured}$=0.014 \citep{bod99} used in Equations \ref{eqn10} and \ref{ratioTest}
with a requirement for $r>1$.  Our previous CHARA Array
example with $(\sigma_{V^2})_{measured}$=0.020 would require from this
approach $\theta \leq 0.20$ mas, which would demand
distant calibration objects beyond its sensitivity limits.
Fortunately techniques have been developed with the apparent ability to predict
stellar angular sizes to better than 10\%, such as spectral energy distribution fitting
(eg. \citet{black1994} and \citet{cohen1999}
agree with interferometric measures at the $\sim$few percent level),
allowing for use of the very long baseline instruments such as the CHARA Array
in an absolute fashion.

Clearly more sophisticated approaches to angular size estimation can
be undertaken for interferometer calibrator stars, presumably with
less susceptibility for size estimation bias, but the blackbody example is illustrative in
how it demonstrates potential bias within an estimation technique.
One of the most useful aspects of an astronomical interferometer,
however, is its ability to mask bias in a calibrator size prediction
technique for a sufficiently unresolved calibration source, and in
doing so, deliver absolutely calibrated visibilities.  Such an approach is
not merely useful but essential to calibrate and verify predictive techniques of
ever-increasing accuracy.

\section{Taylor Series Bias in the Error Propagation Technique}\label{sec_Bias}

The `routine propagation of errors' given in Equation \ref{propOfErrors}
is based upon just the first term of the Taylor series,
which is subject to inaccuracies as the equation becomes more non-linear.
This particular approximation is increasingly inaccurate for non-linear
equations.
Expanding upon our discussion of error propagation in \S \ref{errprop}
to probe the significance of the higher order terms,
we may expand Equation \ref{UDeqn} in a Taylor series about
a given spatial frequency $\mu$:
\begin{equation}
w(x|\mu) = w(\mu | \mu) + (x-\mu)w'(\mu | \mu) + {(x-\mu)^2 \over 2!} w''(\mu | \mu) + ...
\end{equation}
The average of $w(x|\mu)$ can be written as
\begin{equation}\label{wxu_avg}
\overline{w(x|\mu)} = w(\mu | \mu) + {\sigma_x^2 \over 2!} w''(\mu | \mu) + ...
\end{equation}
since, to first order, the $(x-\mu)w'(\mu | \mu)$ term drops out if
$x$ is centered around the mean $\mu$.  The usual error propagation presented
in Equation \ref{propOfErrors} assumes the last term in Equation \ref{wxu_avg} is also negligible, which
represents the Taylor series bias in the error propagation method:
\begin{equation}
bias_{T} = \overline{w(x|\mu)} - w(\mu | \mu) = {\sigma_x^2 \over 2!} w''(\mu | \mu)
\end{equation}
From Equations \ref{UDeqn}, \ref{jincFunction} and \ref{eqn10},
we may write this as
\begin{equation}
bias_{T} =  {\sigma_x^2 \over 2!}
\left(
8 ( jinc'(\mu))^2 - 8 jinc(\mu) jinc''(\mu)
\right)
\end{equation}
Both $jinc$ and $jinc'$ are found in Equation \ref{jincFunction}, and a
derivation of $jinc''$ may be referenced in the Appendix.

\begin{figure}
      \plotone{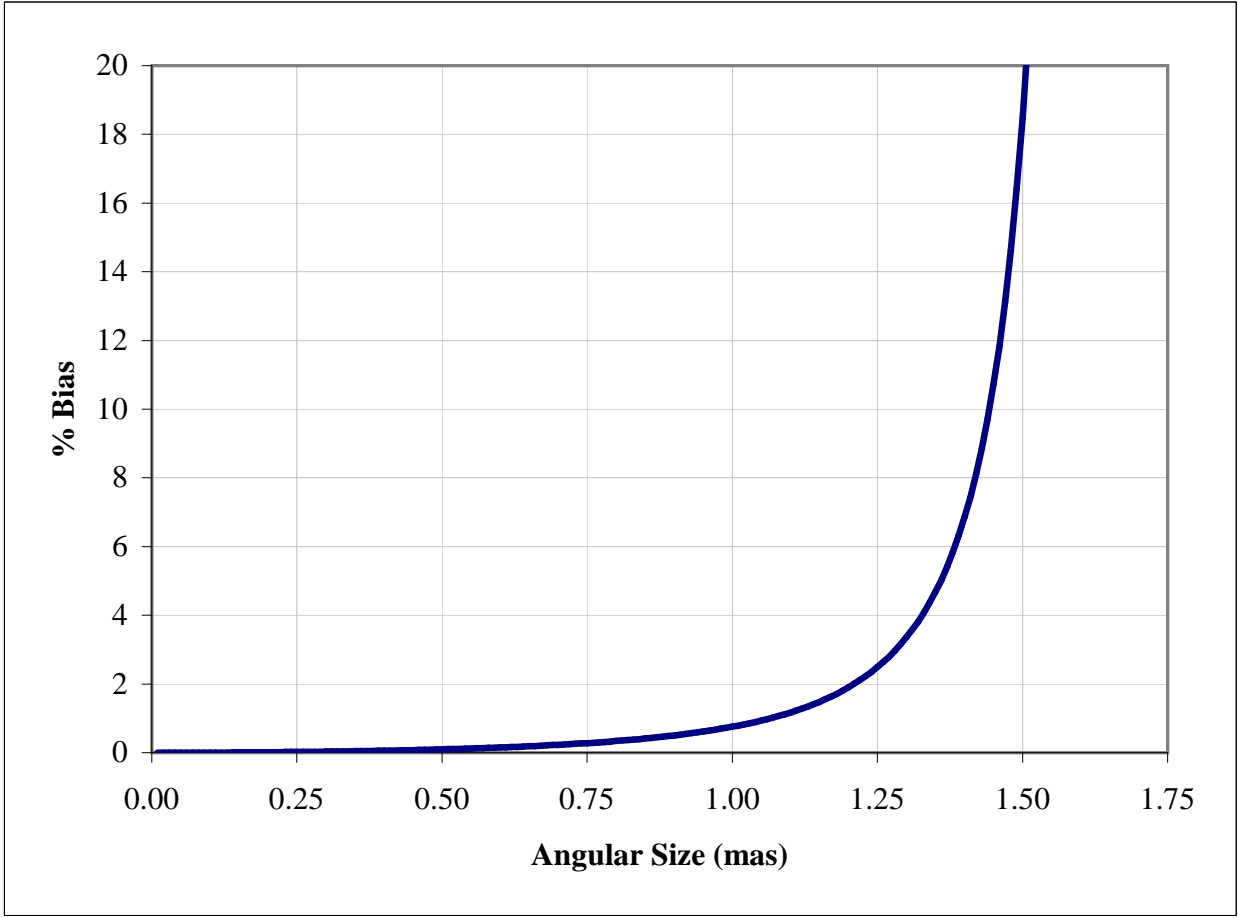}
      \caption{\label{fig_Bias}Percentage Taylor series bias in a $V^2$ measurement
      as discussed in \S \ref{sec_Bias} due to
      calibrator size prediction error of 5\%, as a function of expected calibrator size.}
      \end{figure}

For most applications (including the examples given herein), $\sigma_x$
is dominated by the uncertainty in predicted calibrator angular size,
$(\sigma_\theta)_{predicted}$.  As such, our example 5\% error in $\theta$ means
$\sigma_x = 5\% \times x$; the percentage bias as a function
of calibrator predicted visibility ($bias_T(\mu) / V(\mu)^2$)
is plotted in Figure \ref{fig_Bias} for our CHARA Array test case.
The $bias_T$ term in this case starts to grow exponentially at $\theta\gtrsim 1.25$ mas;
because of this, the `superresolved' calibrator regime indicated in Figure \ref{fig5}
and discussed in \S \ref{sec_ratio_test} is undesirable for use as a source
of calibrators.

\section{Discussion}
Predictive techniques are clearly imperfect - otherwise, why would
we bother with measuring stellar angular sizes through $V^2$
measurements in the first place? As such, it is essential that work
be carried out in the regime that is unaffected by potential bias in
the calibrator angular size predictive technique, or bias from
non-linearities in the visibility function.

Permutations upon the sample CHARA Array case in \S \ref{merit_eval} are worth
considering.  While a 5\% angular size prediction error is
reasonable to expect for most calibration sources, for those sources
with the very best {\it a~priori} spectrophotometric
characterization, a 2.5\% prediction error may be possible.  For
this case, a 0.5\% error in the knowledge of operational wavelength
($\sigma_\lambda=0.01 \mu$m for $K_s$ in Equation \ref{eqn9}) still
only contributes to the  $V^2$ prediction error value by a factor of
approximately $\sim$1.02; prior angular size knowledge at the $<$1\% level is
necessary for this error term to contribute at a level greater than
$\sim$1.10.

As the measurement precision increases (and
$(\sigma_{V^2})_{measured}$ decreases), the unity crossing point seen in
Figure \ref{fig5} (where one passes from the absolute measurement regime into the relative
measurement regime) decreases in value, ranging from $\sim$0.62 mas
for the case of 5\% errors (typically associated with non-spatially
filtered systems), down to $\sim$0.40 mas for 1.5\% errors (typical
of systems with spatial filtering).
This is rather intuitive: as
one's interferometric instrumentation improves in its ability to
precisely measure visibilities, the degree to which that
instrumentation is sensitive to potential biases in calibrator
visibility prediction increases.

These values scale with spatial frequency, which itself will scale linearly
with wavelength and baseline length for individual facilities.  A selection
of currently operational facilities is cited in Table \ref{tab_current_fac}, along with
their relevant operational parameters of operational baseline, wavelength, and
cited measurement precision.  From these values, the maximum calibrator angular size
for absolute angular size measurements is derived using the process
found in \S \ref{sec_ratio_test}, assuming a 5\% angular size estimation error.



\begin{deluxetable}{cccccll}
\tablecolumns{4}
\tablewidth{0pc}
\tablecaption{Maximum calibrator sizes for absolute calibration of $V^2$ measurements
for a variety of current interferometric facilities, assuming 5\% calibrator
size estimate errors.\label{tab_current_fac}}
\tablehead{
\colhead{} & \colhead{Maximum} & \colhead{} & \colhead{Cited V2}
& \colhead{Maximum calibrator}
& \colhead{}
& \colhead{}
\\
\colhead{Facility} & \colhead{Baseline (m)} & \colhead{Band} & \colhead{Meas. Error}
& \colhead{size (mas)}
& \colhead{Notes}
& \colhead{Reference}
 } \startdata
CHARA & 330 & K & 0.04 & 0.77 & No spatial filtering & \citet{van05} \\
CHARA & 330 & K & 0.02 & 0.45 & Spatial filtering & Under development \\
IOTA & 38 & J & 0.046 & 2.3 & No spatial filtering & \citet{mil05} \\
NPOI & 37.5 & V & 0.02 & 1.0 & No spatial filtering & \citet{tyc04} \\
PTI & 110 & K & 0.014 & 1.1 & Spatial filtering & \citet{bod99} \\
VLTI & 187 & K & 0.004 & 0.31 & Spatial filtering,  & \citet{ker03} \\
  &   &   &   &   &  photometric monitoring &   \\
\enddata
\end{deluxetable}



\section{Conclusion}
Given the common use of optical and near-infrared interferometers to establish
basic stellar parameters such as linear radius and effective temperature,
it is of paramount importance to clearly understand the operational regime
of one's instrument as defined not only by its intrinsic capabilities, but
also by the particulars of the observing technique.  As shown in \S \ref{sec_visfn_and_errors},
in the case where resolved
calibrators are being employed, the interferometric visibility measures provided
are relative and as such are subject to biases - known and unknown -
in the calibrator diameter estimation process employed.

The use of
unresolved calibrators found in the regime as defined by the ratio test of
Equation \ref{ratioTest}
is {\it essential} to making absolute measurements.
Additionally,
as seen in \S \ref{sec_Bias}, the non-linear nature of the visibility function
makes the routine propagation of errors incorrect for very low visibilities, and as such,
biases the $V^2$ measurements as well.

\acknowledgments

We would like to thank
Theo ten Brummelaar and Andy Boden for proofreading and thoughtful suggestions.
This manuscript significantly benefited from the comments of an anonymous referee.
Portions of this work were
performed at the California Institute
of Technology under contract with the National Aeronautics and
Space Administration.

\section{Appendix.  Derivation of $jinc''(x)$}\label{jinc''derivation}

Starting with the two following Bessel function identities:
\begin{equation}
{d \over dx}[x^m J_m(x)] = x^m J_{m-1}(x)
\end{equation}
and
\begin{equation}
J_{-m}(x) = (-1)^m J_m(x),
\end{equation}
we may use the recurrence relation
\begin{equation}
J_\nu(z) = {2(\nu-1)\over z} J_{\nu-1}(z)-J_{\nu-2}(z)
\end{equation}
as applied to $J_2$,
\begin{equation}
J_2(z)  = {2\over z} J_{1}(z)-J_{0}(z),
\end{equation}
and explicitly work out $jinc''(x)$:
\begin{equation}
jinc''(x)={d \over dx} jinc'(x) =  {d \over dx} \left( {-J_2(x) \over x} \right)=
{d \over dx} \left[ {1\over x} \left( J_{0}(x)-{2\over x} J_{1}(x) \right) \right]
\end{equation}
\begin{equation}
=  \left[ {-1\over x^2} \left( J_{0}(x)-{2\over x} J_{1}(x) \right) \right]
 + \left[ {1\over x} \left( -J_{1}(x)+{2\over x} J_{2}(x) \right) \right]
\end{equation}


\begin{thebibliography}{}
\bibitem[Airy(1835)]{airy1835} Airy, G.~B. 1835, Trans. Camb. Phil. Soc., 5, 283
\bibitem[Bessell \& Brett(1988)]{besse1988} Bessell, M.~S.~\& Brett, J.~M.\ 1988, \pasp, 100, 1134
\bibitem[Boden et al.(1999)]{bod99} Boden, A.~F., et al.\ 1999, \apj, 515, 356
\bibitem[Born \& Wolf(1980)]{born1980} Born, M.~\& Wolf, E.\ 1980, Oxford: Pergamon Press, 1980, 6th corrected ed.
\bibitem[Blackwell \& Lynas-Gray(1994)]{black1994} Blackwell, D.~E.~\& Lynas-Gray, A.~E.\ 1994, \aap, 282, 899
\bibitem[Bracewell(2000)]{brace2000} Bracewell, R.~N.\ 2000, The Fourier transform and its applications / Ronald N.~Bracewell.~Boston : McGraw Hill, c2000.~(McGraw-Hill series in electrical and computer engineering.~Circuits and systems)
\bibitem[Cohen et al.(1996)]{cohen1996} Cohen, M., Witteborn, F.~C., Carbon, D.~F., Davies, J.~K., Wooden, D.~H., \& Bregman, J.~D.\ 1996, \aj, 112, 2274
\bibitem[Cohen et al.(1999)]{cohen1999} Cohen, M., Walker, R.~G., Carter, B., Hammersley, P., Kidger, M., \& Noguchi, K.\ 1999, \aj, 117, 1864
\bibitem[Colavita(1999)]{colav1999} Colavita, M.~M.\ 1999, \pasp, 111, 111
\bibitem[di Benedetto(1993)]{diben1993} di Benedetto, G.~P.\ 1993, \aap, 270, 315
\bibitem[Dyck, Benson, \& Ridgway(1993)]{dyck1993} Dyck, H.~M., Benson, J.~A., \& Ridgway, S.~T.\ 1993, \pasp, 105, 610
\bibitem[Domiciano de Souza et al.(2003)]{domic2003} Domiciano de Souza, A., Kervella, P., Jankov, S., Abe, L., Vakili, F., di Folco, E., \& Paresce, F.\ 2003, \aap, 407, L47
\bibitem[Dyck, Benson, van Belle, \& Ridgway(1996)]{dyck1996} Dyck, H.~M., Benson, J.~A., van Belle, G.~T., \& Ridgway, S.~T.\ 1996, \aj, 111, 1705
\bibitem[Hajian et al.(1998)]{hajian1998} Hajian, A.~R., et al.\ 1998, \apj, 496, 484
\bibitem[Kervella et al.(2003)]{ker03} Kervella, P., Th{\' e}venin, F., S{\' e}gransan, D., Berthomieu, G., Lopez, B., Morel, P., \& Provost, J.\ 2003, \aap, 404, 1087
\bibitem[Millan-Gabet et al.(2005)]{mil05} Millan-Gabet, R., Pedretti, E., Monnier, J.~D., Schloerb, F.~P., Traub, W.~A., Carleton, N.~P., Lacasse, M.~G., \& Segransan, D.\ 2005, \apj, 620, 961
\bibitem[Mozurkewich et al.(1991)]{mozur1991} Mozurkewich, D., et al.\ 1991, \aj, 101, 2207
\bibitem[ten Brummelaar et al.(2005)]{tenbrum2005} ten Brummelaar, T., et al.\ 2005, \apj, accepted
\bibitem[Tycner et al.(2004)]{tyc04} Tycner, C., et al.\ 2004, \aj, 127, 1194
\bibitem[van Belle(1999)]{vanbe1999} van Belle, G.~T.\ 1999, \pasp, 111, 1515
\bibitem[van Belle et al.(1999)]{vanbe1999b} van Belle, G.~T., et al.\ 1999, \aj, 117, 521
\bibitem[van Belle et al.(2005)]{van05} van Belle, G.~T., et al.\ 2005, \apj, submitted
\end{thebibliography}
\end{document}